\documentclass[twocolumn,aps,prl,showpacs,superscriptaddress]{revtex4}
\usepackage[dvips]{graphicx}
\usepackage{bm}

\begin{document}
\title{Lifetime positron annihilation spectroscopy and photo-inactivated bacteria}

\author{L V Elnikova}

\address{A. I. Alikhanov Institute for Theoretical and Experimental Physics, \\B. Cheremushkinskaya st. 25, Moscow 117218, Russia}

\begin{abstract}
Combined positron and visible light irradiations in the photodynamical therapy (PDT) applications are analyzed. Objectives and goals of PDT are killing or irreversible oxidative damage of pathogenic cells, or rather their cell walls, cell membranes, peptides, and nucleic acids by photo-activated oxygen of photosentizer
injected into the target cell during light irradiation.
In this paper, the arguments for involving of lifetime positron annihilation spectroscopy to control of photodamaging cells in the course of the PDT procedure are given on the examples of gram-positive and gram-negative bacteria basing on a brief survey of the literature.
\end{abstract}
\maketitle
\section{Introduction}
One of the most topical applications in use of affect irradiation onto living organisms might be photodynamical therapy (PDT) combined with positron annihilation methods.

PDT utilizes light of any waveband in combination with a
photosensitizing agent to induce a phototoxic reaction which results with damage or death of pathogenic cells. Antibacterial PDT, directed against bacterial and yield cells, has most demonstrative evidence in dermatology \cite{Jori}. At present, PDT is a main competitor for the antibiotical cure. If penicillin, discovered by Fleming in 1928, was not spread out in medicine,
PDT would not be forgotten at a long date.

Photosensitizing reactions in PDT are the processes, in which absorption of light
by a photosensitizer (or a dye) induces chemical changes in the outer wall at the surface of several types
of bacterial and yeast cells, increases their permeability,
and allows significant amounts of photosensitizer to be
accumulated at the level of the cytoplasmic membrane. Two types of such reactions may carry out, either via radical
mechanism (type I) or energy migration to produce reactive singlet
oxygen (type II). The target cells of bacteria are classified to gram-positive (Gram (+), e.g. murein sacculus) and gram-negative (Gram (-), e.g. Escherichia coli (\textit{E. coli})), they differ in structure and thickness of the peptidoglycan layer.
The positive charge of the dye appears to promote a
tight electrostatic interaction with negatively charged sites
at the outer surface of any species of bacterial cells.

Singlet oxygen $^1$O$_2$ was first observed in 1924. In
1931, Kautsky first proposed that $^1$O$_2$ might be a reaction intermediate in dye-sensitized photooxygenation \cite{ref4ZhaoL}.
Singlet oxygen is metastable states of triplet oxygen (O$_2$) with more high energy is less stable than triplet oxygen (O$_2$).
The energy difference between the lowest energy of singlet state
and the lowest energy of triplet state is about 11400 K ($T_e$ ($a^1\Delta_g \leftarrow X_3\Sigma^-_g$) = 0.98 eV (=94.2 kJ/mol))
and corresponds to the transition in near-infrared at $\sim$ 1270 nm.

The theory of molecular orbital predicts three low-lying excited singlet states of triplet molecular oxygen O$_2(X_3\Sigma^-_g)$: O$_2(a^1\Delta_g)$, O$_2(a'^1\Delta'_g)$, and O$_2(b^1\Sigma^+_g)$,
which differ in only spins and occupation of antibonding degenerated $\pi_g$ orbitals. Amid these states, namely O$_2(a^1\Delta_g)$ is called "singlet oxygen" as non-degenerated and more long-living state. Because of the differences in the electron shells, singlet and triplet oxygen differ in their chemical properties.
Lifetime of singlet oxygen in vacuum equals 72 min and significantly decreases in dependence on a media.
Singlet oxygen reacts with many kinds of biological molecules such as DNA, proteins and lipids \cite{ref5ZhaoL}.

Some notions to develop PAS as a new
noninvasive technique for the detection of molecular damage by UV
radiation have issued in \cite{Jean} and references therein.
The experiments on UV irradiated mouse's \cite{Jean} and cancer-diseased \cite{Jean 3166} skin strikingly demonstrate the perspectives of positron annihilation spectroscopy (PAS) in medicine.

There were observed, that the long-living $o$-Ps is trapping into the 10-15 nm depth of the outer cell layer \cite{Jean}. These scales imply to be comparable to penetration of photoinactivated dye oxygen into the outer layer of gram-negative bacteria, which is of just appropriate to porine and lipopolysaccharide size.

These data motivate the consideration the perspectives of positrons and Ps in the PDT use. To future experimental confirmations and medical applications, we rest here at the evaluation of positronium states for photoirradiated \textit{E. coli} and the porphyrin photosentizer, 4 N--methyl-pyridinium (meso) (T$_4$MPyP), in the presence of the Tris-EDTA agent.

\section{Trial Gram (-) bacterium in the PDT procedure}
Consider the photosensitization of \textit{E. coli} with visible light of 250 W tungsten lamps for T$_4$MPyP due to the PDT action (Figures 1, 2). The presence of the Tris-EDTA agent \cite{Jori}, \cite{Jori3} in the system might be useful to enhancing of the membrane permeability and facilitating the penetration of phototoxic
molecules to the cytoplasmic membrane. The addition of
Tris-EDTA to Gram (-) bacteria removes the divalent cations
(e.g. Ca$^{2+}$, Mg$^{2+}$ ions) which are present in large numbers to stabilize adjacent negative charged lipopolysaccharides molecules at the outer membrane.

\begin{figure}\label{f1}
\includegraphics*[width=70mm]{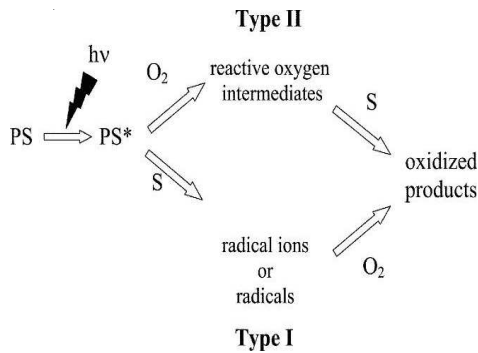}
\caption{\small Pathway of type I and type II reactions of a light absorbing photosensitizer.
After light activating of the ground state of a photosensitizer (PS), activated form of PS* can follow two alternative pathways via reactive singlet oxygen ($^1$O$_2$), hydrogen peroxide, hydroxyl radical (type II) or organic substrate (S) (type I). The intermediates react rapidly with their surroundings: cell wall, cell membrane, peptides, nucleic acids, [1].}
\end{figure}

\begin{figure}\label{f2}
\includegraphics*[width=70mm]{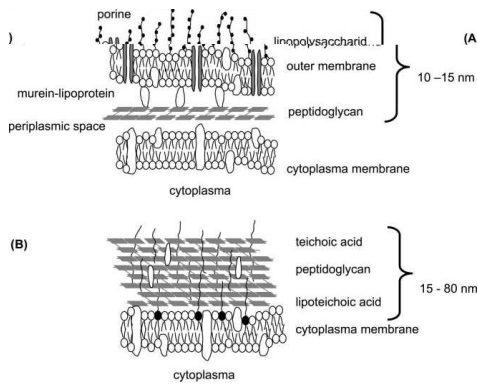}
\caption{\small
Schematic representations of the arrangement of the cell walls of gram-negative (A) and gram-positive (B) bacteria. Gram (-) bacteria cell wall consists of a thin, inner wall composed of 2-3 layers of peptidoglycan (2-3 nm thick), a periplasmic space and an outer lipid bilayer (7 nm). The outer membrane contains phospholipids, lipoproteins, lipopolysaccharides and proteins like porins (A).
Gram (+) bacteria cell wall appears as a 15-80 nm thick cell wall composed of up to 100 interconnecting layers of peptidoglycan (B). Teichoic acids are interwoven in the peptidoglycan layers. Some have a lipid attached (lipoteichoic acid). Also proteins are ingrained in the peptidoglycan layers, [1].}
\end{figure}

Typically, cells were incubated with 8.4 $\mu$M porphyrin solution (1 mL) for 5 min at 37$^o$C, cell pellets were washed once with 5 mM phosphate buffer, pH=7.4, and treated with 2$\%$ aqueous sodium dodecyl sulphate (SDS) in order to disrupt the cells and obtain the incorporation of the porphyrin in a monomeric state into the surfactant micelles.

Absorbed maximum of T$_4$MPyP is lying at 424 nm, the extinction coefficient equals to 194 M$^{-1}$cm$^{-1}$, and respective quantum yield of singlet oxygen is 0.74 \cite{Jori3}.

In the 12 mL'th Pyrex test tube, there are 10$^6$ cell/mL.
The cell survival has controlled due to the standard procedure \cite{Jori3}. During irradiation time 1, 5, and 10 min, 8.4 $\mu$M'th T$_4$MPyP kills 0.0, 3.1, and 4.5 the \textit{E. coli} cells respectively.

\section{Irradiated cells and positronium states}
Biochemical analysis performed on irradiated
cells suggest that the cytoplasmic membrane is an important
target of the photo-process \cite{Jori2}.
To reach this membrane from the outside, singlet oxygen have to destroy an outer cell layer (Figure 2). In particular, the result of its damage should be phase transformations or the pore formation of the outer lipidic layer of a cell. In this respect, DPPC \cite{Jean_lipids} and stratum corneum \cite{gpb} have been studied with the Doppler-broadening and PALS techniques (without specific therapeutic applications).

Consider the main results of photo-destructive reactions of photosentizers in the bacterial layer structure and chemical agents, and the positronium formation after that.

\subsection{Role of singlet oxygen}
Singlet oxygen is essentially used in PTD \cite{Jori}, in the considered scheme, it is produced by meso-Tetra(4-N-methyl-pyridyl)porphin (Figure 3).
T$_4$MPyP is cationic porphyrin charged from +2 to +4. After neutralization of a whole outer layer with chemical or biological agents, one may approximately consider a target cell as neutral before immobilizing a dye (a photosentizer) and light irradiation.

\begin{figure}\label{f3}
\includegraphics*[width=70mm]{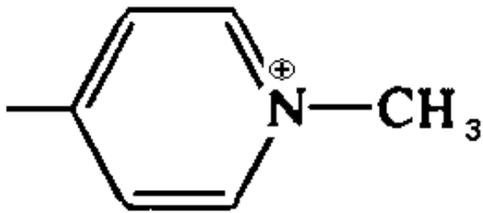}
\caption{\small The chemical structure of meso-Tetra(4-N-methyl-pyridyl)porphin (T$_4$MPyP).
}
\end{figure}
As it has been is known, 
molecular triplet oxygen in the ground state \cite{Gold}, 
\cite{Ferrell} is a quencher for the Ps formation 
due to the mechanism of dissociative trapping, where the energies and the
trapping cross sections are $E_{maxI}=6.2$ eV, $E_{maxII}=8.3$ eV, $\sigma_{maxI}=\sigma_{maxII}=1.3\cdot10^{-18}$
cm$^2$.

Photodynamic
reactions based on primary porphyrin photoreduct ion by the oxidizing
compound are an example of a type I mechanism where molecular oxygen
plays the role of an electron acceptor \cite{Krasnovsky}:
\begin{equation}
\begin{array}{c}
  P+h\nu \longrightarrow ^1P^* \longrightarrow ^3P^*D\longrightarrow ^\bullet P^-+^\bullet D^+ ,\\
  ^\bullet P^*+O_2 \longrightarrow P+ ^\bullet O^-_2,
\end{array}
\end{equation}
where $P$, $^1P^*$ and $^3P^*$ are photosensitizer molecules in the ground, excited
singlet and triplet states, and $D$ is a substrate of photooxygenation. In the
alternative "type II" mechanism, the primary event is the energy transfer from triplet photosensitizers to dioxygen with population of its singlet $^1\triangle_g$ state
($^1O_2$); then, singlet oxygen oxidizes appropriate substrates \cite{Krasnovsky}:
\begin{equation}
\begin{array}{c}
  P+h\nu \longrightarrow ^1P^* \longrightarrow ^3P^*O_2\longrightarrow ^1O_2+P, \\
  ^1O_2 +D \longrightarrow D_{ox}.
\end{array}
\end{equation}

Phosphorescence kinetics after irradiation is described
by the equation \cite{Krasnovsky}
\begin{equation}
L(t)=\frac{k_{gen}[^3P]_0[O_2]}{\tau_r(1/\tau_t-1/\tau_\triangle)}[e^{-t/\tau_\triangle}-e^{-t/\tau_t}],
\end{equation}
where $k_{gen}$ is the rate constant of energy transfer from $^3P^*$ to O$_2$ resulting in
$^1$O$_2$ generation, $\tau_r$ is the $^1$O$_2$ radiative lifetime; $\tau_\triangle$ is the real $^1$O$_2$ lifetime, $\tau_t$ is the $^3P^*$ lifetime; [$^3P]_0$ is the concentration of triplet photosensitizer molecules just after irradiation; [O$_2$] is the dioxygen concentration in solution.

For some agents, there was found (references in  \cite{Krasnovsky}), that the quenching of photosensitized $^1$O$_2$
phosphorescence obeyed
the Stern-Volmer equation, which may be simplified to the next form:
\begin{equation}
\tau_\triangle=1/k_QC_{max},
\end{equation}
where $C_{max}$ the molar concentration of a neat
solvent used as a quencher. In different solvents, $\tau_\triangle$ lyes in the range from 10 to 250 $\mu$s.

For example, under excitation by Nd?Yag laser, 532 nm, photosensitized singlet oxygen phosphorescence in
air-saturated D$_2$O for the tetra-($p$-sulfophenyl) porphyrin photosentizer has the peak achieved in $\sim$ 10 $\mu$s after
the laser flash.

In \cite{Krasnovsky} and references therein, there is proven
phosphorescence emitted by solvated $^1$O$_2$ molecules in H$_2$O, by comparing the $^1$O$_2$ spectra with those in a gas phase.

The quantum yield $\varphi_{ph}$ corresponds to an equation
\begin{equation}
\varphi_{ph}=\varphi_\triangle\varphi_r=\varphi_\triangle t_{\triangle}/t_r,
\end{equation}
where $\varphi_\triangle$ is the quantum yield of $^1$O$_2$ generation by a photosensitizer, $\varphi_r$ is the
$^1$O$_2$ phosphorescence quantum yield, $t_r$ is the $^1$O$_2$ radiative lifetime. The zero-time intensity ($I_0$, in quanta per second) is calculated by extrapolation of the exponential phosphorescence decay to zero time \cite{Krasnovsky}:
\begin{equation}
I_0=I_{las}\varphi_\triangle/t_r,
\end{equation}
where $I_{las}$ is the number of quanta absorbed by a photosensitizer during a laser flash.

The total phosphorescence
yield in photosensitizer solutions is significantly less than 100$\%$. The phosphorescence quantum yields in hydrogen-atom
containing solvents are 0.02 $\%$ and less. The lowest $\varphi_{ph}$ was found in water,
the natural medium of living organisms.

The experimental observation of the $^1$O$_2$ fluorescence in the case of a mouse with implanted cancer tumor loaded with Photofrin II and tetra methyl pyridyl porphin (TMPP, 10$^{-4}$M) \cite{Parker} in water phosphate buffer.  The author of \cite{Parker}, Parker, used the Lexel Model 700L dye laser providing an output wavelength in the range of 600 to 700 nm and fixed at 630 nm, its an average power of was 20 mW. In
the TMPP solution, photosensitized phosphorescence at 1270 nm was
detected. Parker received $\tau_\triangle$=3.2$\mu$s and $\tau_r$=2.4$\mu$s.

Review data, received before 1998 on another agents, photosentizers and target objects one may find once more in \cite{Krasnovsky}.

The rate constants of $^1$O$_2$ interaction with biomolecules can be measured
using chemical $^1$O$_2$ traps (\cite{Foote}, \cite{Lindig}, \cite{Matheson}) or by quenching $^1$O$_2$ phosphorescence (in biological quenchers, such as glycine, tryptophan and so on, $k_q$ is of the order of 10$^7$-10$^8$ (see references in \cite{Krasnovsky})). Rate constants of the $^1$O$_2$ quenching by nucleotides, saccharides and organic acids approximately equal to 10$^4$-10$^7$.

So, the quantum yield of singlet oxygen may serve as a criterion of a cell damage.

From the other hand, the oxygen molecule is a very effective quenching agent for positronium formation \cite{Ferrell},
owing to two unpaired electrons of the ground-state, molecular O$_2$ exhibits both spinflip and nonspin flip quenching.

Paul, Lee and Celitans \cite{Paul}, \cite{Lee}, \cite{Lee2}
concluded, that in a liquid phase, the above mentioned quenched
effect (the decreasing of $\tau_2$ in a positronium spectra) is described by the quenching constant \cite{Gold} 
\begin{equation}\label{1}
\sigma_qv=(\sigma_qv)_{gas}+K_{PsO_2}.
\end{equation}
The first term in the right side of (\ref{1}) means the temperature-independent conversion constant
in a gas phase (the subscript "gas"). Second one denotes the constant for formation of the
PsO$_2$:K$_{PsO_2}$ molecule, which equals to K$^o_{PsO_2}e^{-E/RT}$, where $E$ is an activation energy of a medium molecule, $R$ is the universal gas constant and
$T$ is the absolute temperature.

\subsection{PAL spectra of SDS}
SDS, indispensable in the given PDT process, should be aggregated in a micellar phase due to its phase diagram. In \cite{SDS}, the micellar SDS contribution in PS lifetime spectra was studied. Lifetime spectra were
analyzed in terms of four components ascribed to
$p$-Ps, free e$^+$ and $o$-Ps in the aqueous ($o$-Ps$^{aq}$) and
organic ($o$-Ps$^{org}$) subphases, in increasing order of
the lifetimes $\tau_i$, $i=1,2,3,4$.

Quantitatively, the molar reaction rate constant ($k_{diff}$) is given by Smoluchowsky equation
\begin{equation}
k_{diff}=4\pi DRN_A/1000,
\end{equation}
where $N_A$ is the Avogadro number, $R=R_{Ps}+R_{core}$ and $D=R_{Ps}+R_{mic}$ are the sum of reaction radii and diffusion coefficients of the reactants respectively (see ref. in \cite{SDS}). Each diffusion coefficient may be expressed as a
function of the hydrodynamic radii of the reactants, $R_h$:
$D=\frac{k_BT}{6\pi \eta R_h}$,
where $k_B$ is the Boltzmann constant, $T$ is the absolute
temperature, and $\eta$ is the viscosity of the medium.
The trapping processes from water to organic substrates $o$-Ps$^{org}$ are given as \cite{Dup ref}:
\begin{equation}
2\gamma \stackrel{\lambda^0_3}\longleftarrow o-Ps^{aq}(+micelle) \stackrel{k(t)C_{mic}}\longrightarrow o-Ps^{org}\stackrel {\lambda_4^0}\longrightarrow 2\gamma.
\end{equation}

With no addition of NaCl \cite{Dup ref}, the total intensity associated
to $o$-Ps$^{aq}$ is $I_{tot} = I_3 + I_4 = 22 \pm 0.1$ $\%$. By
comparison with $I^0_3=27$ $\%$ for pure water, this is
due to some slight inhibition
induced by the sulphate polar heads. In the
presence of NaCl, $I_{tot}$ is slightly lower and constant,
at $21.4 \pm 0.1$ $\%$.

The appeared results declared that $\tau_1$ and $\tau_2$ were constant at 178$\pm$21 and 402$\pm$4 ps, respectively.
The value of $\tau_1$ is significantly higher than the
intrinsic $p$-Ps lifetime in vacuum, $\tau_s=$125 ps
($\lambda_s$= 8 ns$^{-1}$, $\lambda_3^0$= 1/1.795 ns $^{-1}$ at 303 K,  $\tau_4$=3.87 $\pm$ 0.05
ns \cite{Dup ref}.

\subsection{Evaluation of spectral contributions of lipopolysaccharides and porins}
Consider the next appropriate results. One of Gram (+) bacteria \cite{gpb}, stratum corneum, was
studied with PALS to qualify the pore diameters.
More specifically, Yucatan miniature pig stratum corneum, separated with heat, was irradiated with polycarbonate (of positron annihilation lifetime 2.103$\pm$0.076 ns) due to sandwich scheme with the $^{22}$Na source 0.51 MeV before samples (15$\times$15$\times$0.5 mm) isolated by the enveloping aluminium foil.

As it has been generally known, the thickness of an outer layer of Gram (+) bacteria is 15-80 nm (Figure 2).
There, pore diameters (from 0.54 to 0.6 nm) in cyclodextrins and polycarbonats were measurable with PALS by means of an application of the well-known Tao-Eldrup formula.

In contrast to stratum corneum, a Gram (-) bacterium has an
outer layer of the 10-15 nm thickness, and shows low level of
photosensitivity, perhaps due to highly organized outer wall,
including murein and a trilamellar structure outside of the
peptidoglycan layer \cite{Jori3}. Therefore, to increase the
permeability of Gram (-) bacteria, the biological agent such
as T$_4$MPyP is involving in PDT procedures.
So, the logarithmic decrease in the overall survival of \textit{E. coli} cells irradiated in the presence of the 8.4$\mu$M T$_4$MPyP porphyrin is 4.5 after 10 min irradiation.

The reader may find the detailed description of cell killing experiments on Gram (-) \textit{E. coli}, for instance, in \cite{Jori3} and \cite{paper}.

There, pores may distinct with cylindrical geometry, and expected to be of the same character sizes, as those in a Gram (+) bacterium. And this fair illustration
of the Gram (+) bacterial PALS application might be spreadable
out onto other bacterial species.

\section{Sum of components of expected spectra of aqueous \textit{E. coli}-- T$_4$MPyP}
It known that the cytoplasm of an organic cell (a main
target of the photo-process \cite{Jori2})
consists of water (75-85$\%$), protein with a mean molecular mass of 10-20$\%$, lipids with a mean molecular mass 2-3$\%$, and other compounds (2$\%$) \cite{Krasnovsky}. Only four amino
acids (tryptophan, histidine, methionine and cysteine) actively interact with singlet oxygen, therefore ? protein macromolecule contains 60 residues of active amino acids. The cytoplasm density is known to be about 1 g/sm$^3$. Molar concentrations of water, proteins and lipids in the cytoplasm are 44, 0.004 and 0.04 M respectively. The mean $k_q$ values for active aminoacids is about $2\cdot10^{7}$M$^{-1}$s$^{-1}$.
So, the total rate of $^1$O$_2$ deactivation has been calculated as follows \cite{Krasnovsky}:
$\sum_g(k_gC_g)=5\cdot10^6$s$^{-1}$,
where $g$ is a component index. Hence, $\tau_\triangle \approx 200 ns$. The quenching activities of protein amino acids greatly exceed those of lipids and water, water contributes 2-3$\%$ into overall quenching. For a cytoplasm membrane $\tau_\triangle \approx 40 ns$ \cite{Krasnovsky}.

The diffusion length of singlet oxygen in different species is estimated as follows
\begin{equation}
l_\triangle \simeq 1/\sqrt{6D\tau_{\triangle}},
\end{equation}
where $D$ is the $^1$O$_2$ diffusion coefficient presented for the discussed medium in \cite{100}. So, in H$_2$O, a lipid membrane and a cytoplasm, $l_\triangle=$1900, 2200 and 90 ${\AA}$ respectively.

Comparison the lifetime spectral contributions of species is shown in Table 1.

\begin{table}[h]
\caption{\label{ex}Contribution of different agents into lifetime spectra}
\begin{center}
\begin{tabular}{llllll}
Medium &$\tau_3$ $o$-Ps, ns & & $\tau_\triangle$, ns $^1$O$_2$\\
water &1.8 \cite{Dup84}& & 1...4 \cite{Krasnovsky}\\
SDS &1.795 \cite{Dup ref}& & -- \\
T$_4$MPyP & -- & & 3200 \cite{Parker} (TMPP) \\
\textit{E.coli} outer layer &--&  & 50 \cite{Krasnovsky}\\
\textit{E.coli} cytoplasmic & -- &  & 40 \cite{Krasnovsky}\\
membrane& & & & & \\
\textit{stratum corneum} cytoplasmic &2.1--2.2 \cite{gpb}&  & $\sim$40 \cite{Krasnovsky} \\
membrane& &  \\
\end{tabular}
\end{center}
\end{table}

\section{Notes on apparatus realization}
There are several features of the PDT-positron combined irradiation.

Typical PALS experiments with living systems (\cite{gpb}, \cite{Uchi} etc) are carried out in frames of the sandwich type configuration of a sample with the positron source (for example, $^{22}$Na). However, such a construction does not allow to study a living system in an equivalent to a living media an aqueous aggregation, and also causes troubles in delivery of light from a laser or a lamp. An alternative bulk construction schemes of a samples in ampules, applicable for liquids on the whole, are described, for instance, in \cite{Dup bulk}.

\section{Conclusions}
The above arguments have proved the possibility of control with positron spectroscopy of states of outer membrane layers in bacterial cells irradiated in PDT. The next conclusions have been made.

1. As the lifetime order of magnitudes for $o$-Ps and $^1$O$_2$ is comparable and different for an each medium (with the exception of water), hence, one may measure the Ps lifetime during the photoinactivation by singlet oxygen during the PDT procedure.

2. The scenario involving the reactions of singlet oxygen of PDT into PALS might be directly tested on the Gram (+) bacterial system \cite{gpb} and spread out onto Gram (-) bacteria.

3. There is announced that for both Gram (+) and Gram (-) bacteria, the quenching effect of singlet oxygen will be impaired depending of the cell death.

\subsection*{Acknowledgments}
The author thanks M. G. Strakhovskaya for her introduction in PDT and joint collaboration. In this text, the review by Lingjie Zhao
"Free Radical and Radiation Biology" (Graduate Program 2001) was used.


\section*{References}
\begin{thebibliography}{99}
\bibitem{Jori} Maisch T, Szeimies R--M, Jori G and Abels C 2004 {\it Photochem. Photobiol. Sci.} {\bf3} 907
\bibitem{ref4ZhaoL}Kearns D R 1971 {\it Chemical Reviews} {\bf 71} 395
\bibitem{ref5ZhaoL}Briviba K, Klotz L O and Sies H 1997
{\it Biol. Chem.} {\bf 378} 1259
\bibitem{Jean}Jean Y C, Chen H, Liu G and Gadzia J E 2007 {\it Radiat. Phys. and Chem.} {\bf76} 70
\bibitem{Jean 3166}  Jean Y C, Li Y, Liu G, Chen H,
 Zhang J and Gadzia J E 2006 {\it Applied Surface Science} {\bf 252} 3166 
\bibitem{Jori3} Villanueva A, Bertaloni G and Jori G 1995 {\it J. Braz. Chem. Soc.}
{\bf 71} 123
\bibitem{Jori2} Bertolini G, Rossi F, Valduga G, Jori G and Van Lier J 2006 {\it FEMS Microbiology Letters}
{\bf 71} 149
\bibitem{Jean_lipids} Jean Y C and Hankock A J 1982 \textit{J. Chem. Phys} {\bf77} 5836
\bibitem{gpb} Itoh Y, Shimazu A, Sadzuka Y, Sonobe T, Itai S 2008
{\it Int. J. of Pharmaceutics} {\bf 358} 91
\bibitem{Krasnovsky} Krasnovsky A A, Jr. 1998 \textit{Membr. Cell Biol.} {\bf12} 665
\bibitem{Parker} Parker J G 1987 {\it IEEE Circuits and Devices ??gazine} 10
\bibitem{Foote}Foote C 1976 {\it Free Radicals in Biology} (New York ? San
Francisco ? London: Academic Press)
\bibitem{Lindig}
Lindig B A and Rodgers M A J 1981 {\it Photochem. Photobiol.} {\bf33} 627
\bibitem{Matheson} Matheson I B C, Etheridge R D, Kratowich N R and Li J 1975 {\it J. Photochem.
Photobiol.} {\bf21} 165
\bibitem{Gold}Goldanskii V I 1968 {\it Fizicheskaya khimia positrona i positronia} (Moscow: Nauka)
\bibitem{Ferrell}Ferrell R A 1958 {\it Phys. Rev.} {\bf 110} 1355 
\bibitem{Paul} Paul D A 1959 {\it Canad. J. Phys.} {\bf 37} 1059
\bibitem{Lee} Lee J, Celitans G J 1965 {\it J. Chem. Phys.} {\bf 42} 437
\bibitem{Lee2} Lee J, Celitans G J 1966 {\it J. Chem. Phys.} {\bf 44} 2506
\bibitem{SDS} Bockstahl F, Pachoud E, Dupl\'{a}tre G and Billard I 2000 {\it Chemical Physics} {\bf 256}  307 
\bibitem{Dup ref} Bockstahl F, Dupl\'{a}tre G 1999 {\it Phys. Chem. Chem. Phys.} {\bf 1} 2767
\bibitem{paper} Minnock A, Vernon D I, Schofield J, Griffiths J, Parish J H and Brown S B 2000{\it Antimicrob Agents Chemother.} {\bf 44} 522
\bibitem{100} Fu Y and Kanofsky J R, Jr. 1995 {\it Photochem. Photobiol.} {\bf62} 692
\bibitem{Dup84} Dupl\'{a}tre G, Talamoni J, Abb\'{e} J Ch and Haessler A 1984 {\it Radiat. Phys. Chem.} {\bf 23} 531
\bibitem{Uchi} Uchiyama Y, Ito K, Li H-L, Ujihara Y, Jean Y C 1996, {\it J. of Radioanalyt. and Nucl. Chem.} {\bf 211} 111
\bibitem{Dup bulk} Dupl\'{a}tre G, Al-Shukri L M and Haessler A 1980 {\it J. of Radioanalytical Chem.}  {\bf 55} 199
\end{thebibliography}
\end{document}